\documentclass[12pt]{article}
\catcode`\@=11
\@addtoreset{equation}{section}

\global\arraycolsep=2pt
\usepackage{mathrsfs,amsbsy,amssymb,latexsym,amsfonts,amsmath,cite,bm}
\usepackage{graphicx,color}
\usepackage{physics}
\usepackage{mathtools}
\DeclareFontFamily{U}{BOONDOX-calo}{\skewchar\font=45 }
\DeclareFontShape{U}{BOONDOX-calo}{m}{n}{
  <-> s*[1.05] BOONDOX-r-calo}{}
\DeclareFontShape{U}{BOONDOX-calo}{b}{n}{
  <-> s*[1.05] BOONDOX-b-calo}{}
\DeclareMathAlphabet{\mathcalboondox}{U}{BOONDOX-calo}{m}{n}
\SetMathAlphabet{\mathcalboondox}{bold}{U}{BOONDOX-calo}{b}{n}
\DeclareMathAlphabet{\mathbcalboondox}{U}{BOONDOX-calo}{b}{n}

\usepackage{enumerate}
\usepackage{comment}

\newcommand{\no}{\nonumber}
\newcommand{\pa}{\partial}

\newcommand{\cL}{\mathcal L}
\newcommand{\cM}{\mathcal M}

\newcommand{\ck}{\mathcalboondox k}

\newcommand{\gP}{{\rm\bf P}}
\newcommand{\gJ}{{\rm\bf J}}

\newcommand{\la}{\lambda}

\newcommand{\llangle}{ \langle\!\langle}
\newcommand{\rrangle}{ \rangle\!\rangle}

\newcommand{\ha}{\hat{a}}

\newcommand{\cha}{\check{a}}

\newcommand{\cc}[1]{\overline{#1}}

\allowdisplaybreaks

\begin{document}

\renewcommand{\thefootnote}{\fnsymbol{footnote}}

\begin{flushright}
KUNS-2878
\end{flushright}
\vspace*{0.5cm}

\begin{center}
{\Large \bf Integrable deformed $T^{1,1}$ sigma models   \\[5pt]
from 4D Chern-Simons theory }
\vspace*{2cm} \\
{\large  Osamu Fukushima$^{\sharp}$\footnote{E-mail:~osamu.f@gauge.scphys.kyoto-u.ac.jp},
Jun-ichi Sakamoto$^{\dagger}$\footnote{E-mail:~sakamoto@ntu.edu.tw},
and Kentaroh Yoshida$^{\sharp}$\footnote{E-mail:~kyoshida@gauge.scphys.kyoto-u.ac.jp}} 
\end{center}

\vspace*{0.4cm}

\begin{center}
$^{\sharp}${\it Department of Physics, Kyoto University, Kyoto 606-8502, Japan.}
\end{center}
\begin{center}
$^{\dagger}${\it Department of Physics and Center for Theoretical Sciences, \\
National Taiwan University, Taipei 10617, Taiwan}
\end{center}

\vspace{1cm}

\begin{abstract} 
Recently, a variety of deformed $T^{1,1}$ manifolds, with which 2D non-linear sigma models (NLSMs) are classically integrable, have been presented by Arutyunov, Bassi and Lacroix (ABL) [arXiv:2010.05573].  We refer to the NLSMs with the integrable deformed $T^{1,1}$ as the ABL model for brevity. Motivated by this progress, we consider deriving the ABL model from a 4D Chern-Simons (CS) theory with 
a meromorphic one-form with four double poles and six simple zeros. We specify boundary conditions in the CS theory that give rise to the ABL model and derive the sigma-model background with target-space metric 
and  anti-symmetric two-form. Finally, we present two simple examples 1) an anisotropic $T^{1,1}$ model and 2) a $G/H$ $\lambda$-model. The latter one can be seen as a one-parameter deformation of the Guadagnini-Martellini-Mintchev model. 
\end{abstract}

\setcounter{footnote}{0}
\setcounter{page}{0}
\thispagestyle{empty}

\newpage

\tableofcontents

\renewcommand\thefootnote{\arabic{footnote}}

\section{Introduction}

A significant subject in String Theory is the integrability in the AdS/CFT correspondence \cite{AdS/CFT1,AdS/CFT2,AdS/CFT3} (For a comprehensive review, see \cite{review}). 
%It may enable us to provide the rigorous proof of AdS/CFT. 
Although there are a lot of research directions, we are interested in the sigma-model classical integrability here. 
In the typical case of AdS/CFT, the string-theory side is basically described by a 2D non-linear sigma model (NLSM)\footnote{We will concentrate on the bosonic part only here.} 
with target space AdS$_5\times$S$^5$ together with the Virasoro constraints after fixing 2D diffeomorphism. 
Then the classical integrability is ensured by the fact that AdS$_5\times$
S$^5$ is described as a symmetric coset which exhibits the $\mathbb{Z}_2$-grading. 

\medskip

Many integrable backgrounds are known apart from AdS spaces and spheres. 
Some of the examples are $\gamma$-deformations of S$^5$ \cite{LM, Frolov}, gravity duals of non-commutative gauge theory \cite{HI,MR} and Schr$\ddot{\rm o}$dinger spacetimes \cite{MMT}. Such integrable backgrounds may be constructed by performing Yang-Baxter deformations \cite{Klimcik1,Klimcik2} of AdS$_5\times$S$^5$ \cite{DMV1,DMV2,KMY1}. There are other integrable-deformation methods such as bi-Yang-Baxter deformations \cite{Klimcik2,Klimcik3} and $\lambda$-deformations \cite{lambda1,lambda2,lambda3,lambda4}. 
It would be worth noting that 2D integrable NLSMs and integrable deformations of them can be described in a unified way based on a 4D Chern-Simons (CS) theory \cite{CY, DLMV} (For related progress, see \cite{FSY1,FSY2,FSY3,Costello:2020lpi,Schmidtt:2019otc,Tian:2020ryu,Chen1,Chen2,Lacroix:2020flf,Benini:2020skc,Caudrelier:2020xtn   }). 

\medskip 

On the other hand, there are a lot of non-integrable backgrounds such as AdS black holes \cite{AdS-BH} 
and AdS solitons \cite{AdS-soliton1, Ishii,AdS-soliton2}.  
The $T^{1,1}$ background \cite{T11}, whose metric is given by 
\begin{eqnarray}
ds^2 = \frac{1}{6}\sum_{r=1}^{2}\left(d\theta_{r}^2+\sin^2\theta_{r}\, d\phi_{r}^2\right) +\frac{1}{9}\big(d\psi + \cos\theta_{1}\, d\phi_{1} + \cos\theta_{2}\, d\phi_{2}\big)\,, 
\end{eqnarray}
is also one of the non-integrable examples \cite{BZ,BZ2}. For the coset construction of $T^{1,1}$ and its Yang-Baxter deformation, see \cite{Crichigno:2014ipa,Rado:2020yhf}. 
The AdS$_5\times T^{1,1}$ geometry is well studied because it is a gravity dual of $\mathcal{N}=1$ superconformal field theory (SCFT) \cite{Klebanov:1998hh}. In relation to SCFT, possible generalizations or deformations of this geometry have intensively been studied. 

\medskip

Recently, Arutyunov, Bassi and Lacroix (ABL) \cite{Arutyunov:2020sdo} have found a family of integrable deformed $T^{1,1}$ NLSMs\footnote{The discussion in \cite{Arutyunov:2020sdo} is based on an affine Gaudin model \cite{Vicedo:2017cge,Delduc:2019bcl,Vicedo:2019dej} and 
covers more general cases. This family of $T^{1,1}$ models is a special case of it. For the off-critical value of the $B$-field, classical chaos appears for some initial conditions \cite{Rigatos:2020hlq,Kushiro}.}. 
We refer to the NLSMs with the integrable deformed $T^{1,1}$ as the ABL model for brevity. 
In this paper, we will derive the ABL model from a 4D CS theory. We start from a certain meromorphic one-form with four double poles and six simple zeros. Then by taking an appropriate boundary condition, we can reproduce 
the classical action of 2D NLSM with four parameters (up to the overall factor). This is nothing but the ABL model. Then we explicitly derive the sigma-model background with target-space metric and anti-symmetric 
two-form. Finally, we present two simple cases 1) an anisotropic $T^{1,1}$ model and 2) a $G/H$ $\lambda$-model. The latter one can be seen as a one-parameter deformation of the Guadagnini-Martellini-Mintchev (GMM) model \cite{Guadagnini:1987ty}.   

\medskip 

This paper is organized as follows.  Section 2 is a short review of a derivation of 2D NLSMs 
from a 4D CS theory. This part is basically based on the seminal work \cite{DLMV}. 
Then in section 3, we derive the ABL model and the sigma-model background is explicitly computed.  
In addition, two simple examples are presented.  
Section 4 is devoted to conclusion and discussion. Appendix A explains a scaling limit of the ABL model which leads to the GMM model by following \cite{Arutyunov:2020sdo}.

\section{2D NLSM from 4D CS theory} 

In this section, we give a derivation of 2D NLSMs from a 4D CS theory by following \cite{CY,DLMV}.

\medskip

Let $G$ be a Lie group with the Lie algebra $\mathfrak{g}$, and $\mathfrak{g}^{\mathbb{C}}$ denotes the complexification of $\mathfrak{g}$\,.
We now consider a 4-dimensional space $\cM\times \mathbb{C}P^1$\,, where $\cM$ and $\mathbb{C}P^1$ are parametrized by coordinates $(\tau,\sigma)$ and $(z,\bar{z})$, respectively.
A 4D CS action is defined as \cite{CY}\footnote{For the notation and convention here, see \cite{FSY1}. },  
\begin{align}
S[A]=\frac{i}{4\pi}\int_{\cM\times\mathbb{C}P^1} \omega\wedge CS(A)\,, 
\label{4dcs}
\end{align}
where $A$ is a $\mathfrak{g}^\mathbb{C}$-valued one-form and $CS(A)$ is the CS three-form defined as 
\begin{align}
CS(A)\equiv \left\langle A,dA+\frac{2}{3}A\wedge A\right\rangle\,. 
\end{align}
$\langle\cdot,\cdot\rangle$ is a non-degenerate adjoint-invariant bilinear form $\mathfrak{g}^{\mathbb{C}}\times\mathfrak{g}^{\mathbb{C}}\to\mathbb{C}$\,.
Then $\omega$ is a meromorphic one-form defined as 
\begin{align}
\omega\equiv \varphi(z)dz \label{omega}
\end{align}
and $\varphi$ is a meromorphic function on $\mathbb{C}P^1$\,. 
This function is found to be a twist function 
characterizing the Poisson structure of the underlying integrable field theory \cite{Vicedo:2019dej}.  
The pole and zero structure of $\varphi$ will be important in the following discussion.  
We denote the sets of poles and zeros by $\mathfrak{p}$ and $\mathfrak{z}$\,, respectively.

\medskip

Note that 
an extra gauge symmetry
\begin{align}
A\mapsto A+\chi\, dz
\label{extra gauge}
\end{align}
can alway gauge away the $z$-component of $A$ like
\begin{align}
A=A_\sigma\, d\sigma+A_\tau\, d\tau+A_{\bar{z}}\, d\bar{z}\,.
\end{align}

\medskip 

By taking a variation of the action (\ref{4dcs})\,, we obtain 
the bulk equation of motion  
\begin{align}
\omega\wedge F(A)=&0\,, \qquad F(A)\equiv dA+A\wedge A
\label{bulk eom}
\end{align}
and the boundary equation of motion 
\begin{align}
d\omega \wedge \langle A,\delta A\rangle=&0\,.\label{boundary eom}
\end{align}
The boundary conditions satisfying (\ref{boundary eom}) play an important role to describe integrable deformations \cite{CY,DLMV}.
Note that the boundary equation of motion (\ref{boundary eom})  does not vanish only on
$\cM\times\mathfrak{p}\subset \cM\times \mathbb{C}P^1$\,, because the relation
\begin{align}
d \omega = \partial_{\bar{z}}\varphi(z)\, d\bar{z} \wedge dz 
\label{d-omega-delta}
\end{align}
indicates that only the pole of $\varphi$ can contribute as a distribution. 
This can be seen by rewriting the equation (\ref{boundary eom}) to
% and (\ref{d-omega-delta}) 
\begin{align}
\sum_{x\in\mathfrak{p}}\sum_{p\geq0}\left(\operatorname{res}_x \xi_x^p \omega\right)\epsilon^{ij}\frac{1}{p!}\partial_{\xi_x}^p
\langle A_i,\delta A_j \rangle\big|_{\cM \times \{x\}}=0\,,\label{general boundary}
\end{align}
where $\epsilon^{ij}$ is the antisymmetric tensor.
Here the local holomorphic coordinates $\xi_x$ are defined as $\xi_x\equiv z-x$ 
for $x\in\mathfrak{p}\backslash\{\infty\}$ and $\xi_\infty\equiv1/z$ if $\mathfrak{p}$ 
includes the point at infinity. The expression (\ref{general boundary}) manifestly shows that the boundary equation 
of motion has the support only on  $\cM\times\mathfrak{p}$\,.

\medskip

In terms of the components, the bulk equation of motion (\ref{bulk eom}) reads
\begin{align}
\partial_\sigma A_\tau -\partial_\tau A_\sigma+[A_\sigma,A_\tau]=&0\,,\\
\omega\,\left(\partial_{\bar{z}} A_\sigma -\partial_\sigma A_{\bar{z}}+[A_{\bar{z}},A_\sigma]
\right)=&0\,,\\
\omega\,\left(\partial_{\bar{z}} A_\tau -\partial_\tau A_{\bar{z}}+[A_{\bar{z}},A_\tau]
\right)=&0\,.
\end{align}
The factor $\omega$ is kept since $\partial_{\bar{z}}A_\sigma$ 
and $\partial_{\bar{z}}A_\tau$ are in general distributions on $\mathbb{C}P^1$ supported by $\mathfrak{z}$\,.

\subsection*{Lax form}

By performing a formal gauge transformation
\begin{align}
A=-d\hat{g}\hat{g}^{-1}+\hat{g}\mathcal{L}\hat{g}^{-1} 
\label{L def}
\end{align}
with a smooth function $\hat{g}:\cM\times\mathbb{C}P^1
\rightarrow G^{\mathbb{C}}$, 
the $\bar{z}$-components of $\cL$ can be taken to zero:
\begin{align}
\mathcal{L}_{\bar{z}}=0\,.  
\label{gauge fix}
\end{align}
Hence the one-form $\mathcal{L}$ takes the form
\begin{align}
\mathcal{L}\equiv\mathcal{L}_{\sigma}d\sigma+\mathcal{L}_{\tau}d\tau\,.
\end{align}
The one-form will be specified as a Lax pair for 2D theory later, and so we refer to $\mathcal{L}$ as the Lax form.

\medskip 

The bulk equations of motion (\ref{bulk eom}) in terms of the Lax form $\mathcal{L}$ are expressed as
\begin{align}
\partial_\tau \mathcal{L}_\sigma -\partial_\sigma \mathcal{L}_\tau+&[\mathcal{L}_\tau,
\mathcal{L}_\sigma]=0\,,\\
\omega\wedge \partial_{\bar{z}}\mathcal{L}&=0\,.\label{L holomorphic}
\end{align}
These equations means
 that $\mathcal{L}$ is a meromorphic one-form with poles at the zeros of $\omega$\,,  
namely $\mathfrak{z}$ can be regarded as the set of poles of $\mathcal{L}$\,. 

\subsection*{Reality condition}

To ensure the reality of the 4D action (\ref{4dcs}) and the resulting action (\ref{2d action}), we suppose some condition for the form of $\omega$ and the configuration of $A$ \cite{DLMV}.

\medskip 

For a complex coordinate $z$, an involution $\mu_{\mathrm{t}}:\mathbb{C}P^{1}\to\mathbb{C}P^{1}$ is defined by complex conjugation $z\mapsto\bar{z}$\,.
Let $\tau:\mathfrak{g}^{\mathbb{C}}\to\mathfrak{g}^{\mathbb{C}}$ be an anti-linear involution which satisfies
\begin{align}
\cc{\langle B,C \rangle} = \langle \tau B, \tau C \rangle\,, \qquad ^{\forall}B,C\in \mathfrak{g}^{\mathbb{C}}\,.
\end{align}
Then a real Lie subalgebra $\mathfrak{g}$ of $\mathfrak{g}^{\mathbb{C}}$ is given as the set of the fixed points under $\tau$\,.
The associated operation to the Lie group $G$ is denoted by $\tilde{\tau}: G^{\mathbb{C}}\to G^{\mathbb{C}}$\,.

\medskip

One can see that the action (\ref{4dcs}) is real if $\omega$ and $A$ satisfy
\begin{align}
\cc{\omega}=&\mu_{\mathrm{t}}^{*}\omega\,,\label{reality omega}  \\
\tau A=&\mu_{\mathrm{t}}^{*}A\,.\label{reality A}  
\end{align}
Recalling the relation (\ref{L def}), the condition
\begin{align}
 \tilde{\tau} \hat{g}=\mu_{\mathrm{t}}^{*}\hat{g}\,,\qquad & \tau \mathcal{L}=\mu_{\mathrm{t}}^{*}\mathcal{L}\,,\label{reality L}
\end{align}
leads to (\ref{reality A}).

\subsection*{From 4D to 2D via the archipelago conditions}

The 4D action (\ref{4dcs}) can be reduced to a 2D action with the WZ term when $\hat{g}$ satisfies the archipelago condition\cite{DLMV}.
By performing an integral over $\mathbb{C}P^1$\,, we obtain
%We consider a reduction of the 4D action (\ref{4dcs}) into a 2D action with the WZ term.
%For up to second-order poles of the one-form $\omega$, such a sufficient condition is given as the archipelago condition \cite{DLMV}.
%When $\hat{g}$ satisfies the archipelago condition, the 4D action (\ref{4dcs}) is reduced to
\begin{align}
S\left[\left\{g_{x}\right\}_{x \in \mathfrak{p}}\right]=\frac{1}{2} \sum_{x \in \mathfrak{p}} \int_{\cM}\left\langle\operatorname{res}_{x} (\varphi \,\mathcal{L}), g_{x}^{-1} d g_{x}\right\rangle
-\frac{1}{2} \sum_{x \in \mathfrak{p}}\left(\operatorname{res}_{x} \omega\right) \int_{\cM\times[0,R_x]} I_{\mathrm{WZ}}\left[g_{x}\right]\,.
\label{2d action}
\end{align}
%by performing an integral over $\mathbb{C}P^1$\,.
Here $I_{WZ}[u]$ is the Wess-Zumino (WZ) three-form defined as
\begin{align}
I_{WZ}[u]\equiv \frac{1}{3}\langle u^{-1}du,u^{-1}du\wedge u^{-1}du\rangle\,,
\end{align}
where $R_x$ is the radius of the open disk on $\mathbb{C}P^1$\,.

\medskip

The action (\ref{2d action}) is invariant under a gauge transformation 
\begin{align}
g_x\mapsto g_x h\,,\qquad \mathcal{L}\mapsto h^{-1}\mathcal{L}h+h^{-1}dh\,, 
\label{2d-gauge}
\end{align}
with a local function $h:\cM\rightarrow G$\,. 
%Since $h$ depends only on $(\tau,\sigma)$\,, this invariance is regarded as a 2D gauge symmetry.
One can seen this as the residual gauge symmetry after taking the gauge (\ref{gauge fix}).
%Note here that we have not imposed the reality condition by following \cite{FSY1}, in comparison to \cite{DLMV}. 
%The reality condition will be introduced later when fixing a boundary condition of $\hat{g}$\,. 

\section{The ABL model from 4D CS theory}

In this section, we shall consider 2D NLSMs with a family of deformed $T^{1,1}$ manifolds, which have been presented by 
Arutyunov-Bassi-Lacroix \cite{Arutyunov:2020sdo}. We will refer to them as the ABL model for brevity as explained in Introduction.  

\medskip 

Here, let us reproduce the ABL model from 4D CS theory.
In the ABL model,  the 2D surface $\cM$ is embedded into the Lie group $G\times G$\,. 
By defining a subgroup $H\subset G$ as fixed points of an involutive automorphism, this model exhibits a gauge $H_{\rm diag}$-symmetry, where  
$H_{\rm diag}$ is the diagonal subgroup of $G\times G$\,. Then the phase space is reduced to a coset $(G\times G)/H_{\rm diag}$\,.
%The Lie groups $G_1$ and $G_2$ are associated with the Lie algebras $\mathfrak{g_{1}}, \mathfrak{g}_2$\,, respectively.
%Thus the 1-form $A$ now takes value in $\mathfrak{g}^{\mathbb{C}}=\mathfrak{g}_{1}^{\mathbb{C}}\oplus\mathfrak{g}_{2}^{\mathbb{C}}$\,.

\subsubsection*{Twist function}

Let us start with the following meromorphic one-form,   
\begin{align}
    \omega=\varphi_{\rm ABL}(z) \,dz=2K\frac{z(z^2-\zeta_+^2)(z^2-\zeta_-^2)}{\prod_{i=1}^2(z^2-z_i^2)^2}dz\,,\label{eq:twist-new}
\end{align}
where $\varphi_{\rm ABL}(z)$ is a twist function with $\zeta_{\pm}\in\mathbb{C}P^1$ and $z_{1},z_{2}\in\mathbb{R}$\,.
This $\omega$ has the four double poles and the six simple zeros
\begin{align}
    \mathfrak{p}=\{\pm z_1, \pm z_2\}\,,\qquad \mathfrak{z}=\{0, \pm \zeta_+\,,\pm \zeta_-,\infty\}\,.
\end{align}
The twist function in (\ref{eq:twist-new}) corresponds to the case with $N=2$ and $T=2$ in (3.14) in \cite{Arutyunov:2020sdo}.

%\medskip
%
%For convenience of the later discussion, we will refer to the first $G$ in $G\times G$ as $G_1$ and the second one as $G_2$\,.
%Moreover, the associated Lie algebras are denoted by $\mathfrak{g}_1\,, \mathfrak{g}_2$\,, respectively.

\subsubsection*{Boundary condition}

In specifying a 2D integrable model associated with $\omega$\,, we need to choose a solution 
to the boundary equations of motion, 
\begin{align}
   \epsilon^{\alpha\beta} \llangle(A_{\alpha},\partial_{\xi_{p}} A_{\alpha}),\delta(A_{\beta}, \partial_{\xi_{p}} A_{\beta})\rrangle_{p}=0\,,\qquad p\in\mathfrak{p}\,.
   \label{eq:beom}
\end{align}
Here the double bracket is defined as
\begin{align}
    \llangle (x,y), (x',y')\rrangle_{p}&\equiv(\text{res}_{p}\,\omega)\langle x, x'\rangle+(\text{res}_{p}\,\xi_{p}\omega)\left(\langle x, y'\rangle+\langle x', y\right\rangle)\no\\
    &=c_p\left(\langle x, y'\rangle+\langle x', y\rangle\right)\,,
    \label{def:inner-double}
\end{align}
where the overall constants $c_p\,(p\in \mathfrak{p})$ are given by
\begin{align}
    c_{\pm z_1}=\pm \frac{K(z_1^2-\zeta_1^2)(z_1^2-\zeta_2^2)}{2z_1(z_1^2-z_2^2)^2}\,,\qquad c_{\pm z_2}=\pm \frac{K(z_2^2-\zeta_1^2)(z_2^2-\zeta_2^2)}{2z_2(z_1^2-z_2^2)^2}\,.
\end{align}
The boundary equations of motion (\ref{eq:beom}) take the same form as in the PCM case.

\medskip

To derive the ABL model, we take the following solution: 
\begin{align}
(A|_{z=\pm z_1},\partial_{z} A|_{z=\pm z_1})\in \{0\}\ltimes\mathfrak{g}_{{\rm ab}}\,,\qquad (A|_{z=\pm z_2},\partial_{z} A|_{z=\pm z_2})\in \{0\}\ltimes\mathfrak{g}_{{\rm ab}} \,,
\label{eq:A-bsol}
\end{align}
where $\mathfrak{g}_{\rm ab}$ is an abelian copy of $\mathfrak{g}$\,.

\subsubsection*{Lax form}

Before deriving the sigma model action, we shall introduce our notation used in the following.
We take $\hat{g}$ at each pole of the twist function (\ref{eq:twist-new}) as
\begin{align}
\begin{split}
    \hat{g}(\tau,\sigma,z)|_{z=z_1}&=g_1(\tau,\sigma)\,,\qquad \hat{g}(\tau,\sigma,z)|_{z=-z_1}=\tilde{g}_1(\tau,\sigma)\,,\\
    \hat{g}(\tau,\sigma,z)|_{z=z_2}&=g_2(\tau,\sigma)\,,\qquad \hat{g}(\tau,\sigma,z)|_{z=-z_2}=\tilde{g}_2(\tau,\sigma)
    \,,
    \end{split}
\end{align}
where $g_{k}\in G\:(k=1,\dots,4)$\,.
Note that $g_k$ take values in $G$ (not $G^{\mathbb{C}}$) due to the reality condition (\ref{reality L})\,.
The associated left-invariant currents are defined as
\begin{align}
    j_1& \equiv g_1^{-1}dg_1\,,\qquad \tilde{j}_1 \equiv \tilde{g}_1^{-1}d\tilde{g}_1\,,\qquad
    j_2 \equiv g_2^{-1}dg_2\,,\qquad \tilde{j}_2 \equiv \tilde{g}_2^{-1}d\tilde{g}_2
    \,,
\end{align}
and the relations between the gauge field $A$ and the Lax form $\cL$ at each pole are written as
\begin{align}
\begin{split}
    A|_{z=z_1}=-dg_1 g^{-1}_1&+{\rm Ad}_{g_1}\cL|_{z=z_1}\,,\qquad
    A|_{z=-z_1}=-d\tilde{g}_1\tilde{g}_1^{-1}+{\rm Ad}_{\tilde{g}_1}\cL|_{z=-z_1}\,,\\
    A|_{z=z_2}=-dg_2 g^{-1}_2&+{\rm Ad}_{g_2}\cL|_{z=z_2}\,,\qquad
    A|_{z=-z_2}=-d\tilde{g}_2\tilde{g}_2^{-1}+{\rm Ad}_{\tilde{g}_2}\cL|_{z=-z_2}\,,
    \label{eq:A-Lax-new}
    \end{split}
\end{align}
where the adjoint action $\operatorname{Ad}_g:\mathfrak{g}^{\mathbb{C}}\to\mathfrak{g}^{\mathbb{C}}$ is defined as $\operatorname{Ad}_{g}\cL=g\cL g^{-1}$\,.
Taking account of the configuration of the zeros (\ref{eq:twist-new})\,,
we suppose an ansatz for the Lax form as
\begin{align}
    \cL=\left(U_{+}^{[1]}+z\,U^{[2]}_{+}+\frac{U^{[3]}_{+}}{z-\zeta_+}+\frac{U^{[4]}_{+}}{z+\zeta_+}\right)d\sigma^++\left(U_{-}^{[1]}+z^{-1}\,U^{[2]}_{-}+\frac{U^{[3]}_{-}}{z-\zeta_-}+\frac{U^{[4]}_{-}}{z+\zeta_-}\right)d\sigma^{-}\,,\label{eq:Lax-ansatz}
\end{align}
where $U_{\pm}^{[k]}\,(k=1,\dots,4)$ are undetermined smooth functions of $\tau$ and $\sigma$\,, and
the light-cone coordinates are defined as
\begin{align}
\sigma^{\pm} \equiv \frac{1}{2}\left(\tau\pm\sigma\right)\,.
\end{align} 
%As we will see, the ansatz (\ref{eq:Lax-ansatz}) of the Lax form works well for \magenta{two classes of boundary conditions.}

\subsection{2D action}

Let us derive the 2D action by employing the boundary condition (\ref{eq:A-bsol})\,.  

\medskip 

Under the boundary condition (\ref{eq:A-bsol}), the relations in (\ref{eq:A-Lax-new}) are rewritten as 
\begin{align}
    j_{1,\pm}&=U_{\pm}^{[1]}+z_{1}^{\pm1}\,U^{[2]}_{\pm}+\frac{U^{[3]}_{\pm}}{z_1-\zeta_{\pm}}+\frac{U^{[4]}_{\pm}}{z_1+\zeta_{\pm}}\,,\\
    \tilde{j}_{1,\pm}&=U_{\pm}^{[1]}-z_{1}^{\pm1}\,U^{[2]}_{\pm}-\frac{U^{[3]}_{\pm}}{z_1+\zeta_{\pm}}-\frac{U^{[4]}_{\pm}}{z_1-\zeta_{\pm}}\,,\\
     j_{2,\pm}&=U_{\pm}^{[1]}+z_{2}^{\pm1}\,U^{[2]}_{\pm}+\frac{U^{[3]}_{\pm}}{z_2-\zeta_{\pm}}+\frac{U^{[4]}_{\pm}}{z_2+\zeta_{\pm}}\,,\\
    \tilde{j}_{2,\pm}&=U_{\pm}^{[1]}-z_{2}^{\pm1}\,U^{[2]}_{\pm}-\frac{U^{[3]}_{\pm}}{z_2+\zeta_{\pm}}-\frac{U^{[4]}_{\pm}}{z_2-\zeta_{\pm}}\,.   
\end{align}
By solving these equations with respect to $U_{\pm}^{[k]}\,(k=1,2,3,4)$\,, we obtain
\begin{align}
    U_{\pm}^{[1]}&=\frac{(j_{1,\pm}+\tilde{j}_{1,\pm})\left(z_1^2-\zeta_{\pm}^2\right) -(j_{2,\pm}+\tilde{j}_{2,\pm}) \left(z_2^2-\zeta_{\pm}^2\right)}{2 \left(z_1^2-z_2^2\right)}\,,\\
    U_{\pm}^{[2]}&=\frac{z_1^{\mp1}(j_{1,\pm}-\tilde{j}_{1,\pm}) \left(z_1^{\pm2}-\zeta_{\pm}^{\pm2}\right)-z_2^{\mp1} (j_{2,\pm}-\tilde{j}_{2,\pm}) \left(z_2^{\pm2}-\zeta_{\pm}^{\pm2}\right)}{2  \left(z_1^{\pm2}-z_2^{\pm2}\right)}\,,\\
    U_{\pm}^{[3]}&=-\frac{(z_1 z_2)^{\mp1} \left(z_1^2-\zeta_{\pm}^2\right) \left(z_2^2-\zeta_{\pm}^2\right)}{4 \zeta_{\pm} \left(z_1^2-z_2^2\right)}\biggl[z_2^{\pm1}\left(j_{1,\pm} \left(z_1^{\pm1}+\zeta_{\pm}^{\pm1}\right)+\tilde{j}_{1,\pm} \left(z_1^{\pm1}-\zeta_{\pm}^{\pm1}\right)\right)\no\\ 
    &\qquad\qquad\qquad-z_1^{\pm1}\left(j_{2,\pm} \left(z_2^{\pm1}+\zeta_{\pm}^{\pm1}\right)+\tilde{j}_{2,\pm} \left(z_2^{\pm1}-\zeta_{\pm}^{\pm1}\right)\right)\biggr]\,,\\
    U_{\pm}^{[4]}&=\frac{\left(z_1^2-\zeta_{\pm}^2\right) \left(z_2^2-\zeta_{\pm}^2\right)}{4\zeta_{\pm} \left(z_1^2-z_2^2\right) }
    \biggl[z_1^{\mp1}  \left(j_{1,\pm} \left(z_1^{\pm1}-\zeta_{\pm}^{\pm1}\right)+\tilde{j}_{1,\pm} \left(z_1^{\pm1}+\zeta_{\pm}^{\pm1}\right)\right)\no\\
    &\qquad\qquad -z_2^{\mp1} \left(j_{2,\pm} \left(z_2^{\pm1}-\zeta_{\pm}^{\pm1}\right)+\tilde{j}_{2,\pm} \left(z_2^{\pm1}+\zeta_{\pm}^{\pm1}\right)\right)\biggr]\,. 
\end{align}
Then the Lax pair can be rewritten as
\begin{align}
    \cL_{\pm}(z)=\eta_{1,\pm}^{(0)}(z)J^{(0)}_{1,\pm}+\eta_{1,\pm}^{(1)}(z)J^{(1)}_{1,\pm}+\eta_{2,\pm}^{(0)}(z) J^{(0)}_{2,\pm}+\eta_{2,\pm}^{(1)}(z)J^{(1)}_{2,\pm}\,, \label{eq:Lax}
\end{align}
where $J^{(k)}_{s,\pm}~(k=0,1,\, s=1,2)$ are defined as
\begin{align}
\begin{split}
    J^{(0)}_{1,\pm}&=\frac{j_{1,\pm}+\tilde{j}_{1,\pm}}{2}\,,\qquad  J^{(1)}_{1,\pm}=\frac{j_{1,\pm}-\tilde{j}_{1,\pm}}{2}\,,\\
    J^{(0)}_{2,\pm}&=\frac{j_{2,\pm}+\tilde{j}_{2,\pm}}{2}\,,\qquad J^{(1)}_{2,\pm}=\frac{j_{2,\pm}-\tilde{j}_{2,\pm}}{2}\,,
\end{split}
\end{align}
and the coefficients $\eta^{(k)}_{s,\pm}~(k=0,1,\, s=1,2)$ are  
\begin{align}
\begin{split}
   \eta_{\pm,1}^{(0)}(z)&=\frac{(z^2-z_2^2)(z^2_1-\zeta^2_{\pm})}{(z^2-\zeta^2_{\pm})(z^2_1-z_2^2)}\,,\qquad \eta_{\pm,1}^{(1)}(z)=\left(\frac{z}{z_1}\right)^{\pm1}\eta_{\pm,1}^{(0)} \,,\\
   \eta_{\pm,2}^{(0)}(z)&=-\frac{(z^2-z_1^2)(z^2_2-\zeta^2_{\pm})}{(z^2-\zeta^2_{\pm})(z^2_1-z_2^2)}\,,\qquad \eta_{\pm,2}^{(1)}(z)=\left(\frac{z}{z_2}\right)^{\pm1}\eta_{\pm,2}^{(0)} \,.
\end{split}
\end{align}

\medskip

Next, let us evaluate the residues of $\varphi_{\rm ABL}\, \cL$ at $z=\pm z_1\,,\pm z_2$\,.
By using the expression (\ref{eq:Lax}) of the Lax form, we obtain
%\begin{align}
%\begin{split}
 % &  \text{res}_{z=\pm z_{1}}(\varphi_{\rm ABL}\, \cL)  \\
%=&\mp\frac{K}{2 z_1 \left(z_1^2-z_2^2\right)^2}\biggl[\frac{z_1^2-\zeta_{-}^2}{z_1^2-\zeta_{+}^2} \left(-U_{+}^{[2]} \left(z_1^2- \zeta_{+}^2\right)^2+U_{+}^{[3]} (\zeta_{+}\pm z_1)^2+U_{+}^{[4]} (z_1\mp \zeta_{+})^2\right)d\sigma^+\\
%    &\quad+\frac{ z_1^2-\zeta_{+}^2}{z_1^2-\zeta_{-}^2} \left(U_{-}^{[2]}z_1^{-2} \left(z_1^2-\zeta_{-}^2\right)^2+U_{-}^{[3]} (\zeta_{-}\pm z_1)^2+U_{-}^{[4]} (z_1\mp\zeta_{-})^2\right)d\sigma^- \biggr]\,,\\
% &       \text{res}_{z=\pm z_{2}}(\varphi_{\rm ABL}\, \cL) \\ 
%=&\mp\frac{K}{2 z_2 \left(z_1^2-z_2^2\right)^2}\biggl[\frac{z_2^2-\zeta_-^2}{z_2^2-\zeta_+^2}\left(-U_{+}^{[2]} \left(z_2^2-\zeta_+^2\right)^2+U_{+}^{[3]} (\zeta_+\pm z_2)^2+U_{+}^{[4]} (z_2\mp\zeta_+)^2\right)d\sigma^+\\
%        &\qquad +\frac{z_2^2-\zeta_+^2}{z_2^2-\zeta_-^2} \left(U_{-}^{[2]}z_2^{-2} \left(z_2^2-\zeta_-^2\right)^2+U_{-}^{[3]} (\zeta_-\pm z_2)^2+U_{-}^{[4]} (z_2\mp\zeta_-)^2\right)d\sigma^-
%        \biggr]\,,
 %       \label{eq:phiL-res}
 %       \end{split}
%\end{align}
%or equivalently,
\begin{align}
\begin{split}
\operatorname{res}_{z=\pm z_1}(\varphi_{\rm ABL}\cL)=&\,
\left(
-J_{1,+}^{(0)}\rho_{12}^{(0)} + J_{2,+}^{(0)}\rho_{21}^{(0)} \pm J_{1,+}^{(1)}c_{1,+}^{(1)} \pm J_{2,+}^{(1)}\rho_{21}^{(1)}
\right)d\sigma^{+}\\
&+\left(
J_{1,-}^{(0)}\rho_{21}^{(0)} - J_{2,-}^{(0)}\rho_{12}^{(0)} \mp J_{1,-}^{(1)}c_{1,-}^{(1)} \mp J_{2,-}^{(1)}\rho_{12}^{(1)}
\right)d\sigma^{-}\,,\\[3pt]
%%%%%%%%%%%%%%%%%%%%%%%%%%%
\operatorname{res}_{z=\pm z_2}(\varphi_{\rm ABL}\cL)=&\,
\left(
J_{1,+}^{(0)}\rho_{12}^{(0)} - J_{2,+}^{(0)}\rho_{21}^{(0)} \pm J_{1,+}^{(1)}\rho_{12}^{(1)} \mp J_{2,+}^{(1)}c_{2,+}^{(1)}
\right)d\sigma^{+}\\
&+\left(
-J_{1,-}^{(0)}\rho_{21}^{(0)} + J_{2,-}^{(0)}\rho_{12}^{(0)} \mp J_{1,-}^{(1)}\rho_{21}^{(1)} \pm J_{2,-}^{(1)}c_{2,-}^{(1)}
\right)d\sigma^{-}\,,
%%%%%%%%%%%%%%%%%%%%%%%%%%%
  %  \text{res}_{z=\pm z_{1}}(\varphi_{c}\, \cL)&=\left(\pm J_{2,+}^{(1)}\rho_{21}^{(1)}+(-J_{1,+}^{(0)}+J_{2,+}^{(0)})\rho_{21}^{(0)}\pm J_{1,+}^{(1)}c_{1,+}^{(1)}\right)d\sigma^+\\
    %&\quad+\left(\mp J_{2,-}^{(1)}\rho_{12}^{(1)}-(-J_{1,-}^{(0)}+J_{2,-}^{(0)})\rho_{12}^{(0)}\mp J_{1,-}^{(1)}c_{1,-}^{(1)}\right)d\sigma^- \,,\\
       % \text{res}_{z=\pm \red{z_{2}}}(\varphi_{c}\, \cL)&=\left(\mp J_{2,+}^{(1)}c_{2,+}^{(1)}-(-J_{1,+}^{(0)}+J_{2,+}^{(0)})\rho_{12}^{(0)}\pm J_{1,+}^{(1)}\rho_{12}^{(1)}\right)d\sigma^+\\
    %&\quad+\left(\pm J_{2,-}^{(1)}c_{2,-}^{(1)}+(-J_{1,-}^{(0)}+J_{2,-}^{(0)})\rho_{21}^{(0)}\mp J_{1,-}^{(1)}\rho_{21}^{(1)}\right)d\sigma^-\,.
        \label{eq:phiL-res2}
        \end{split}
\end{align}
where the constants $\rho_{rs,\pm}^{(k)}(r,s=1,2\,,k=0,1)$ are defined as 
\begin{align}\begin{split}
    \rho_{11}^{(0)}&=\rho_{22}^{(0)}=\frac{K}{2}\frac{\zeta_-^2-\zeta_+^2}{(z_1^2-z_2^2)^2}\,,\qquad  
    \rho_{12}^{(0)}=K\frac{(z_1^2-\zeta_+^2)(z_2^2-\zeta_-^2)}{(z_1^2-z_2^2)^3}\,,\\ \rho_{21}^{(0)}&=-K\frac{(z_1^2-\zeta^2_-)(z_2^2-\zeta_+^2)}{(z_1^2-z_2^2)^3}\,,
\end{split} \label{rho-para0}
\end{align}
and
\begin{align}\begin{split}
    \rho^{(1)}_{11}&=\frac{K}{2}\frac{(z^4_1-2\zeta^2_+z^2_1+\zeta^2_-\zeta_+^2)}{z_1^2(z_1^2-z_2^2)^2}\,,\qquad 
    \rho_{12}^{(1)}=K\frac{z_2(z_1^2-\zeta_+^2)(z_2^2-\zeta_-^2)}{z_1(z_1^2-z_2^2)^3}\,,\\
    \rho^{(1)}_{21}&=-K\frac{z_1(z_1^2-\zeta^2_-)(z_2^2-\zeta^2_+)}{z_2(z_1^2-z_2^2)^3}\,,\qquad \rho^{(1)}_{22}=\frac{K}{2}\frac{(z_2^4-2\zeta^2_+z_2^2+\zeta^2_-\zeta^2_+)}{z_2^2(z_1^2-z_2^2)^2}\,.
\end{split} \label{rho-para1}
\end{align}
Furthermore, the constants $c_{s,\pm}^{(1)}~(s=1,2)$ are
\begin{align}\begin{split}
    c_{1,-}^{(1)}&=\frac{K\left(z_1^2-\zeta_-^2\right)}{2 z_1^2 \left(z_1^2-z_2^2\right)^3}\left(\zeta_+^2 \left(z_2^2-3 z_1^2\right)+z_1^2 \left(z_1^2+z_2^2\right)\right)\,,\\
     c_{1,+}^{(1)}&=\frac{K \left(z_1^2-\zeta_+^2\right)}{2 z_1^2 \left(z_1^2-z_2^2\right)^3} \left(z_1^2 \left(z_1^2-3 z_2^2\right)+\zeta_-^2 \left(z_1^2+z_2^2\right)\right)\,,\\
    c_{2,-}^{(1)}&=\frac{K \left(z_2^2-\zeta_-^2\right)}{2 z_2^2 \left(z_1^2-z_2^2\right)^3} \left(\zeta_+^2 \left(z_1^2-3 z_2^2\right)+z_2^2 \left(z_1^2+z_2^2\right)\right)\,,\\
     c_{2,+}^{(1)}&=\frac{K \left(z_2^2-\zeta_+^2\right)}{2 z_2^2 \left(z_1^2-z_2^2\right)^3} \left(z_2^2 \left(z_2^2-3 z_1^2\right)+\zeta_-^2 \left(z_1^2+z_2^2\right)\right)\,.  
\end{split}\end{align}
Note that the above constants satisfy the relations
\begin{align}
   \frac{\rho^{(0)}_{12}+\rho^{(0)}_{21}}{2}=-\rho^{(0)}_{11}=-\rho^{(0)}_{22}\,,\qquad  \frac{c_{1,+}^{(1)}+c_{1,-}^{(1)}}{2}=\rho^{(1)}_{11}\,,\qquad 
     \frac{c_{2,+}^{(1)}+c_{2,-}^{(1)}}{2}=-\rho^{(1)}_{22}\,.\label{crho-rel}
\end{align}
By using (\ref{crho-rel}), we obtain
\begin{align}
  & \left\langle\text{res}_{z=z_{1}}(\varphi_{\rm ABL}\, \cL), j_{1}\right\rangle+\left\langle\text{res}_{z=-z_{1}}(\varphi_{\rm ABL}\, \cL),\tilde{j}_{1}\right\rangle\no\\
   =&2\sum_{k=0}^1\left( 2\rho_{11}^{(k)}\left\langle J_{1,+}^{(k)},J_{1,-}^{(k)}\right\rangle+\rho_{12}^{(k)}\left\langle J_{1,+}^{(k)},J_{2,-}^{(k)}\right\rangle+\rho_{21}^{(k)}\left\langle J_{2,+}^{(k)},J_{1,-}^{(k)}\right\rangle\right)d\sigma^+\wedge d\sigma^-\,,\\
   & \left\langle\text{res}_{z=z_{2}}(\varphi_{\rm ABL}\, \cL),j_{2}\right\rangle+\left\langle\text{res}_{z=-z_{2}}(\varphi_{\rm ABL}\, \cL),\tilde{j}_{2}\right\rangle\no\\
   =& 2\sum_{k=0}^1\left(2\rho_{22}^{(k)}\left\langle J_{2,+}^{(k)},J_{2,-}^{(k)}\right\rangle+\rho_{12}^{(k)}\left\langle J_{1,+}^{(k)},J_{2,-}^{(k)}\right\rangle+\rho_{21}^{(k)}\left\langle J_{2,+}^{(k)},J_{1,-}^{(k)}\right\rangle\right)d\sigma^+\wedge d\sigma^-\,.
\end{align}
The residues of $\omega$ at each pole are
\begin{align}
    -\text{res}_{z=\pm z_{1}}\omega=\text{res}_{z=\pm z_{2}}\omega=
    \ck\equiv K\frac{2z_1^2z_2^2+2\zeta^2_-\zeta^2_+-(z_1^2+z_2^2)(\zeta_-^2+\zeta^2_+)}{(z_1^2-z_2^2)^3}\,.
    \label{level-definition}
\end{align}
Then, the $2$D action is given by 
\begin{align}
    S[g_k]&=\int_{\cM}\sum_{r,s=1}^2\left(\rho_{rs}^{(0)}\left\langle J_{r,+}^{(0)},J_{s,-}^{(0)}\right\rangle+\rho_{rs}^{(1)}\left\langle J_{r,+}^{(1)},J_{s,-}^{(1)}\right\rangle \right)2 \,d\sigma^+\wedge d\sigma^-\no\\
    &\qquad +\frac{\ck}{2} \int_{\cM\times[0,R_r]} \big(I_{\mathrm{WZ}}\left[g_{1}\right]+I_{\mathrm{WZ}}\left[\tilde{g}_{1}\right]\big)
-\frac{\ck}{2} \int_{\cM\times[0,R_s]} \big(I_{\mathrm{WZ}}\left[g_{2}\right]+I_{\mathrm{WZ}}\left[\tilde{g}_{2}\right]\big)\,.
    \label{eq:sym-action-lr}
\end{align}

\medskip

Here we would like to impose a relation between $j_s$ and $\tilde{j}_s~(s=1,2)$\,. 
Note that the resulting action (\ref{eq:sym-action-lr}) is invariant under the exchange of $j_1$ and $\tilde{j}_1$ ($j_2$ and $\tilde{j}_2$)\,. 
This invariance is respected if $j_{s},\tilde{j}_{s}$ are related as $\tilde{j}_1=f_1(j_1)$\,, $\tilde{j}_{2}=f_2(j_2)$ with involutive automorphisms $f_s:\mathfrak{g}\to \mathfrak{g}~(s=1,2)$\,.
The maps $f_{s}$ thus satisfy
\begin{align}
    f_s([\mathrm{x},\mathrm{y}])=[f_s(\mathrm{x}),f_s(\mathrm{y})]\,,\qquad f_s\circ f_s(\mathrm{x})=\mathrm{x}\,,\qquad \mathrm{x} \in \mathfrak{g}\,.
\end{align}
By utilizing the involutions $f_s$\,, the vector space $\mathfrak{g}$ 
can be decomposed as {$\mathfrak{g}=\mathfrak{h}\oplus\mathfrak{m}$\,, i.e., the generators 
\begin{align}
   \mathfrak{h}= \langle \gJ_{\ha} \rangle\,, 
\qquad 
\mathfrak{m}= \langle \gP_{\cha}\rangle \,,
\qquad
\ha=1,\dots, \operatorname{dim}\mathfrak{h}\,,\quad \cha=1,\dots, \operatorname{dim}\mathfrak{m}\,,
\end{align}
are introduced so that
\begin{align}
f_s(\gP_{\cha})=-\gP_{\cha}\,,\qquad f_s(\gJ_{\ha})=\gJ_{\ha}\,. 
\label{eq:sym-map}
\end{align}
The vector subspace $\mathfrak{h}$ is also a subalgebra of $\mathfrak{g}$\,, and thus there exists the associated Lie subgroup $H$\,.
Then the projection operators into $\mathfrak{h},\mathfrak{m}$ are defined as
\begin{align} 
P_{(0)}~:~\mathfrak{g} \to \mathfrak{h}\,, \qquad 
P_{(1)}~:~\mathfrak{g}\to \mathfrak{m} \,,
\end{align}
and then $\tilde{j}_s$ and $J_{s}^{(k)}$ are expressed as
\begin{align}
   \tilde{j}_s=f_s(j_s)&=f_s\left(P_{(0)}(j_s)+P_{(1)}(j_s)\right)=P_{(0)}(j_s)-P_{(1)}(j_s)\,, 
    \label{tg-conf}\\
    J_{s}^{(0)}&=P_{s}^{(0)}(j_{s})\,,\qquad J_{s}^{(1)}=P_{s}^{(1)}(j_{s})\,.
\end{align}   
By using the commutation relation of the Lie algebra for the symmetric coset, we can see
\begin{align}
    \langle P_{(0)}(g_s^{-1}dg_s),P_{(0)}(g_s^{-1}dg_s)\wedge P_{(1)}(g_s^{-1}dg_s)\rangle&=0\,,\\
        \langle P_{(1)}(g_s^{-1}dg_s),P_{(1)}(g_s^{-1}dg_s)\wedge P_{(1)}(g_s^{-1}dg_s)\rangle&=0\,.
\end{align}
Hence, we obtain
\begin{align}
    I_{\mathrm{WZ}}\left[g_{1}\right]=I_{\mathrm{WZ}}\left[\tilde{g}_{1}\right]\,,\qquad  I_{\mathrm{WZ}}\left[g_{2}\right]=I_{\mathrm{WZ}}\left[\tilde{g}_{2}\right]\,.
\end{align}
Then, by using the expressions of $\tilde{j}_s$ in (\ref{tg-conf}), the 2D action can be further rewritten as 
\begin{align}
    S[g_1,g_2]=\sum_{r,s=1}^2\int d^2\sigma\,\left(\rho^{(0)}_{rs}\,\left\langle J_{r,+}^{(0)},J_{s,-}^{(0)}\right\rangle+\rho^{(1)}_{rs}\,\left\langle J_{r,+}^{(1)},J_{s,-}^{(1)}\right\rangle  \right)+\ck\,I_{WZ}[g_1]-\ck\,I_{WZ}[g_3]\,.
    \label{ABL-action}
\end{align}
The Lax form (\ref{eq:Lax}) becomes
\begin{align}
    \cL_{\pm}(z)=\sum_{r=1}^2\sum_{k=0}^1\eta_{\pm,r}^{(k)}(z)\,J^{(k)}_{r,\pm}\,. 
\label{ABL-Lax}
\end{align}
The expressions (\ref{ABL-action}) and (\ref{ABL-Lax}) are the same as 
the classical action and the associated Lax pair given in \cite{Arutyunov:2020sdo}. 

\subsubsection*{Gauge invariance}
The action (\ref{ABL-action}) exhibits a local $H_{\rm diag}$-symmetry, which is regarded as a gauge symmetry.
The diagonal subgroup 
$H_{\rm diag}=\{(h,h)\in G\times G\,|\, h\in H\}$
acts on $G\times G$ as
\begin{align}
g_{1}\mapsto g_1h\,,\qquad g_{2}\mapsto g_{2}h\,,
\label{coset-gauge}
\end{align}
where $h$ is a smooth map $\cM\to H$\,.
Noting that the Wess-Zumino terms vary according to the Polyakov-Wigmann formula \cite{Polyakov:1983tt}, we can see that the action (\ref{ABL-action}) is invariant if the following conditions hold:
\begin{align}
&\left\{\begin{array}{l}
\sum_{r,s}^{1,2}\rho_{rs}^{(0)}=0\,,\\
\rho_{11}^{(0)} + \rho_{12}^{(0)} - \frac{\ck}{2} = \rho_{21}^{(0)} + \rho_{22}^{(0)} + \frac{\ck}{2}=0\,,\\
\rho_{11}^{(0)} + \rho_{21}^{(0)} + \frac{\ck}{2} = \rho_{12}^{(0)} + \rho_{22}^{(0)} - \frac{\ck}{2}=0\,,
\end{array}\right.\\
\Leftrightarrow\quad&\rho_{11}^{(0)}=\rho_{22}^{(0)}\,,\quad \rho_{12}^{(0)}-\rho_{21}^{(0)}=\ck\,,\quad 
\frac{\rho_{12}^{(0)}+\rho_{21}^{(0)}}{2}+\rho_{11}^{(0)}=0\,.
\label{coefficient-relation}
\end{align}
These relations are indeed satisfied by the parametrization (\ref{rho-para1}) and (\ref{level-definition})\,.
The gauge invariance under (\ref{coset-gauge}) is nothing but the unbroken part of the 2D gauge invariance under (\ref{2d-gauge})\,. 
Although the original gauge group is $G_{\rm diag}=\{(g,g)\in G\times G\,|\,g\in G\}$\,, the grading condition (\ref{tg-conf}) explicitly break the non-diagonal part 
$G_{\rm diag}/H_{\rm diag}$ part and only the diagonal part $H_{\rm diag}$ survives.

\subsection{Examples}\label{ex}

The resulting action (\ref{ABL-action}) and Lax form (\ref{ABL-Lax}) are a bit abstract 
and complicated. Hence, it is instructive to see a simple case with $G=SU(2)$ and $H=U(1)$\,. Then it is easy to read off the background metric and $B$-field. 

\medskip

The generators of $\mathfrak{su}(2)$ are represented by $\{i\sigma_{a}/2\,,\,i=1,2,3\}$\,, where $\sigma_{a}$ are the Pauli matrices.
The bilinear form $\langle\cdot,\cdot\rangle$ becomes the trace operation.
In this case, the involutive automorphisms $f_{s}$ $(s=1,2)$ are defined as 
\begin{align}
f_{s}\left(\frac{i\sigma_{1}}{2}\right)=-\frac{i\sigma_{1}}{2}\,,\qquad f_{s}\left(\frac{i\sigma_{2}}{2}\right)=-\frac{i\sigma_{2}}{2}\,,
\qquad f_{s}\left(\frac{i\sigma_{3}}{2}\right)=\frac{i\sigma_{3}}{2}\,.
\end{align}
We choose the parameters $\{\phi_{1},\theta_{1},\psi_{1},\phi_{2},\theta_{2},\psi_{2}\}=\{x^\mu\}\,(\mu=1,\dots,6)$ to express $(g_{1},g_{2})\in SU(2)\times SU(2)$ as
\begin{align}\begin{split}
&g_{1}=\exp(\frac{i\sigma_{3}}{2}\phi_{1})\exp(\frac{i\sigma_{2}}{2}\theta_{1})\exp(\frac{i\sigma_{3}}{2}\psi_{1})\,,\\
&g_{2}=\exp(-\frac{i\sigma_{3}}{2}\phi_{2})\exp(-\frac{i\sigma_{2}}{2}\theta_{2})\exp(-\frac{i\sigma_{3}}{2}\psi_{2})\,. 
\end{split}\label{T11-parametrization}
\end{align}
Then the gauge transformation (\ref{coset-gauge}) corresponds to the shift 
\begin{eqnarray}
(\psi_{1},\psi_{2})\mapsto (\psi_{1}+\alpha,\psi_{2}-\alpha)\,. 
\end{eqnarray} 

\subsubsection*{The ABL background}

By substituting the parametrization (\ref{T11-parametrization}), the resulting action is given by
\begin{align}
S[x^{\mu}]=-\frac{1}{4}\int d^2\sigma \left(G_{\mu\nu}+B_{\mu\nu}\right)\pa_{-}x^{\mu}\pa_{+}x^{\nu}\,,
\label{action-metric}
\end{align}
where $G_{\mu\nu}$ and $B_{\mu\nu}$ are the background metric and $B$-field, respectively.
By using the relations (\ref{coefficient-relation}),
$G_{\mu\nu}$ and $B_{\mu\nu}$ are expressed as, respectively, 
\begin{align}\begin{split}
& \frac{1}{2}G_{\mu\nu}dy^{\mu}dy^{\nu} \\ 
=&\,
\sum_{r=1}^{2}\Big[\rho_{rr}^{(1)}\left(d\theta_{r}^2+\sin^2\theta_{r}d\phi_{r}^2\right)
+\rho_{rr}^{(0)}\left(\cos\theta_{r}d\phi_{r} + d\psi_{r}\right)^2\Big]\\
&-\left(\rho_{12}^{(0)}+\rho_{21}^{(0)}\right)\!\big(\cos\theta_{1}d\phi_{1}+d\psi_{1}\big)\big(\cos\theta_{2}d\phi_{2}+d\psi_{2}\big)\\
&-\left(\rho_{12}^{(1)}+\rho_{21}^{(1)}\right)\!\Big[-\sin\theta_{1}\sin\theta_{2}\cos(\psi_{1}+\psi_{2})d\phi_{1}d\phi_{2}
+\cos(\psi_{1}+\psi_{2})d\theta_{1}d\theta_{2}\\
&\hspace{84pt} 
+\sin\theta_{1}\sin(\psi_{1}+\psi_{2})d\phi_{1}d\theta_{2}
+ \sin\theta_{2}\sin(\psi_{1}+\psi_{2})d\theta_{1}d\phi_{2}
\Big]
\end{split}\no\\
\begin{split}
=&\,
\sum_{r=1}^{2}\rho_{rr}^{(1)}\left(d\theta_{r}^2+\sin^2\theta_{r}d\phi_{r}^2\right)
+\rho_{11}^{(0)}\big(d\psi_{1}+d\psi_{2}+\cos\theta_{1}d\phi_{1}+\cos\theta_{2}d\phi_{2}\big)^2\\
&-\left(\rho_{12}^{(1)}+\rho_{21}^{(1)}\right)\!\Big[-\sin\theta_{1}\sin\theta_{2}\cos(\psi_{1}+\psi_{2})d\phi_{1}d\phi_{2}
+\cos(\psi_{1}+\psi_{2})d\theta_{1}d\theta_{2}\\
&\hspace{84pt} +\sin\theta_{1}\sin(\psi_{1}+\psi_{2})d\phi_{1}d\theta_{2}
+ \sin\theta_{2}\sin(\psi_{1}+\psi_{2})d\theta_{1}d\phi_{2}
\Big]\,, 
\end{split}
\end{align}
\begin{align}
B=&\,\frac{1}{2}B_{\mu\nu}dx^{\mu}\wedge dx^{\nu}\no\\
\begin{split}
=&\,
\ck\cos\theta_{1}\,d\phi_{1}\wedge d\psi_{1} + \ck\cos\theta_{2}\,d\psi_{2}\wedge d\phi_{2}\\
&+\left(\rho_{12}^{(0)}-\rho_{21}^{(0)}\right)\big(\cos\theta_{1}d\phi_{1}+d\psi_{1}\big)\wedge\big(\cos\theta_{2}d\phi_{2}+d\psi_{2}\big)\\
&+\left(\rho_{12}^{(1)}-\rho_{21}^{(1)}\right)\Big[
-\sin\theta_{1}\sin\theta_{2}\cos(\psi_{1}+\psi_{2})\,d\phi_{1}\wedge d\phi_{2}
+\cos(\psi_{1}+\psi_{2})\,d\theta_{1}\wedge d\theta_{2}\\
&\hspace{86pt} +\sin\theta_{1}\sin(\psi_{1}+\psi_{2})\,d\phi_{1}\wedge d\theta_{2} + \sin\theta_{2}\sin(\psi_{1}+\psi_{2})\,d\theta_{1}\wedge d\phi_{2}\Big]
\end{split}\no\\
=&\,
\ck\big(d\psi_{1}+d\psi_{2}+\cos\theta_{1}d\phi_{1}\big)\wedge\big(d\psi_{1}+d\psi_{2}+\cos\theta_{2}d\phi_{2}\big)\no\\
&+\left(\rho_{12}^{(1)}-\rho_{21}^{(1)}\right)\Big[
-\sin\theta_{1}\sin\theta_{2}\cos(\psi_{1}+\psi_{2})\,d\phi_{1}\wedge d\phi_{2}
+\cos(\psi_{1}+\psi_{2})\,d\theta_{1}\wedge d\theta_{2}\\
&\hspace{86pt} +\sin\theta_{1}\sin(\psi_{1}+\psi_{2})\,d\phi_{1}\wedge d\theta_{2} + \sin\theta_{2}\sin(\psi_{1}+\psi_{2})\,d\theta_{1}\wedge d\phi_{2}\Big]\,. \no
\end{align}
By taking a gauge choice $\psi_{2}=0$\,, these are further simplified as 
\begin{align}
\begin{split}
\frac{1}{2}G_{\mu\nu}dx^{\mu}dx^{\nu}=&\,
\sum_{r=1}^{2}\rho_{rr}^{(1)}\left(d\theta_{r}^2+\sin^2\theta_{r}d\phi_{r}^2\right)
+\rho_{11}^{(0)}\left(d\psi + \cos\theta_{1}d\phi_{1} + \cos\theta_{2}d\phi_{2}\right)^2\\
&-\left(\rho_{12}^{(1)}+\rho_{21}^{(1)}\right)\Big[-\sin\theta_{1}\sin\theta_{2}\cos\psi \,d\phi_{1}d\phi_{2} + \cos\psi \,d\theta_{1}d\theta_{2}\\
&\hspace{86pt} + \sin\theta_{1}\sin\psi \,d\phi_{1}d\theta_{2} + \sin\theta_{2}\sin\psi\, d\theta_{1}d\phi_{2} \Big]\,,
\end{split}\\
\begin{split}
B=&\,\frac{1}{2}B_{\mu\nu}dx^{\mu}\wedge dx^{\nu}\\
=&\,
\ck\left(d\psi + \cos\theta_{1}d\phi_{1}\right)\wedge \left(d\psi + \cos\theta_{2}d\phi_{2}\right)\\
&+\left(\rho_{12}^{(1)}-\rho_{21}^{(1)}\right)
\Big[ -\sin\theta_{1}\sin\theta_{2}\cos\psi \,d\phi_{1}\wedge d\phi_{2} + \cos\psi\, d\theta_{1}\wedge d\theta_{2} \\
&\hspace{86pt}+ \sin\theta_{1}\sin\psi \,d\phi_{1}\wedge d\theta_{2} + \sin\theta_{2}\sin\psi\, d\theta_{1}\wedge d\phi_{2}\Big]\,.
\end{split}
\end{align}

\medskip 

In the following, we will consider two specific cases by taking some parametrization.

%%%%%%%%%%%%%%%%%%%%%%%%%%%%%%%%%%%%%%%%%%%%%%%%%%%%%%%%%%%%%%

\subsubsection*{EX. 1) Anisotropic $T^{1,1}$ model }

For simplicity, let us first impose the following condition: 
%We notice that this metric is the one of the $T^{1,1}$ manifolds if 
\begin{align}
\rho_{12}^{(1)}+\rho_{21}^{(1)}=0\,.
\label{T11-para}
\end{align}
This condition (\ref{T11-para}) is solved in terms of $z_1,\,z_{2}$ as
\begin{align}
\zeta_{+}^2=\frac{z_{1}^2z_{2}^2}{z_{1}^2+z_{2}^2-\zeta_{-}^2}\quad \Leftrightarrow\quad
\zeta_{-}^2=\frac{-z_{1}^2z_{2}^2+z_{1}^2\zeta_{+}^2+z_{2}^2\zeta_{+}^2}{\zeta_{+}^2}\,.
\end{align}
By introducing a new quantity $r$ defined as 
\begin{align}
r\equiv \rho_{22}^{(1)}/\rho_{11}^{(1)}=\frac{-z_{2}^2+\zeta_{-}^2}{z_{1}^2-\zeta_{-}^2}\,,
\end{align}
the coefficients are expressed as 
\begin{align}\begin{split}
\rho_{11}^{(1)}=&\, \frac{K}{2}\frac{1}{z_{1}^2+rz_{2}^2}\,,\\
\rho_{22}^{(1)}=&\, r\rho_{11}^{(1)}\,,\\
\rho_{11}^{(0)}=&\, \frac{r}{1+r}\rho_{11}^{(1)}\,,\\
\rho_{12}^{(1)}-\rho_{21}^{(1)}=&\, -\frac{4rz_{1}z_{2}}{(1+r)(z_{1}^2-z_{2}^2)}\rho_{11}^{(1)}\,,\\
\ck=&\, -\frac{2r(z_{1}^2+z_{2}^2)}{(1+r)(z_{1}^2-z_{2}^2)}\rho_{11}^{(1)}\,. 
\end{split} \end{align}
Then the metric, $B$-field and the twist function are given by, respectively, 
\begin{align}
\begin{split}
\frac{1}{2}G_{\mu\nu}dx^{\mu}dx^{\nu}=&\,
\rho_{11}^{(1)}\Big[\big(d\theta_{1}^2+\sin^2\theta_{1}d\phi_{1}^2\big)+r\big(d\theta_{2}^2+\sin^2\theta_{2}d\phi_{2}^2\big)\\
&\hspace{25pt}+\frac{r}{1+r}\big(d\psi+\cos\theta_{1}d\phi_{1}+\cos\theta_{2}d\phi_{2}\big)^2\Big]\,,
\end{split}\\
\begin{split}
\frac{1}{2}B_{\mu\nu}dx^{\mu}\wedge dx^{\nu}=&\,
\rho_{11}^{(1)}\Big[-\frac{2r(z_{1}^2+z_{2}^2)}{(1+r)(z_{1}^2-z_{2}^2)}
\left(d\psi + \cos\theta_{1}d\phi_{1}\right)\wedge \left(d\psi + \cos\theta_{2}d\phi_{2}\right)\\
&\hspace{25pt}\pm-\frac{4rz_{1}z_{2}}{(1+r)(z_{1}^2-z_{2}^2)}\\
&\hspace{40pt}\Big( -\sin\theta_{1}\sin\theta_{2}\cos\psi \,d\phi_{1}\wedge d\phi_{2} + \cos\psi\, d\theta_{1}\wedge d\theta_{2} \\
&\hspace{47pt}+ \sin\theta_{1}\sin\psi \,d\phi_{1}\wedge d\theta_{2} + \sin\theta_{2}\sin\psi\, d\theta_{1}\wedge d\phi_{2}\Big)\Big]\,,
\end{split}\\
\varphi_{\rm ABL}=&\,
2K\frac{z\big(rz_{1}^2+z_{2}^2-(1+r)z^2\big)\big((1+r)z_{1}^2z_{2}^2-(z_{1}^2+rz_{2}^2)z^2\big)}{(1+r)(z_{1}^2+rz_{2}^2)(z^2-z_{1}^2)^2(z^2-z_{2}^2)^2}\,.
\end{align}
There remain three independent parameters $\rho_{11}^{(1)},\, r$ 
and $z_{1}/z_{2}$ now. Note here that the original $T^{1,1}$ case is not included in this example because the vanishing $B$-field means that $r=0$\,. The GMM model is also not included. The parameter $r$ may be rather seen as an anisotropic parameter. In the isotropic case with $r=1$\,, the coefficient of the $U(1)$-fiber is fixed as $1/2$ and the $B$-field remains complicated. 

\medskip 

Finally, the Lax pair is given by
\begin{align}\begin{split}
\cL_{+} 
=&\,\frac{1}{(1+r)z_{1}^2z_{2}^2-(z_{1}^2+rz_{2}^2)z^2} \times \\ 
& \qquad \qquad \quad \times \left[z_{1}^2(z_{2}^2-z^2)\left(J_{1,+}^{(0)}+\frac{z}{z_{1}}J_{1,+}^{(1)}\right)+
rz_{2}^2(z_{1}^2-z^2)\left(J_{2,+}^{(0)}+\frac{z}{z_{2}}J_{2,+}^{(1)}\right)
\right]\,,\\
\cL_{-} 
=&\frac{1}{rz_{1}^2+z_{2}^2-(1+r)z^2}\left[(z_{2}^2-z^2)\left(J_{1,-}^{(0)}+\frac{z_{1}}{z}J_{1,-}^{(1)}\right)
+r(z_{1}^2-z^2)\left(J_{2,-}^{(0)}+\frac{z_{2}}{z}J_{2,-}^{(1)}\right)
\right]\,.
\end{split}\end{align}

\subsubsection*{EX. 2) $G/H$ $\lambda$-model}

As the second example, let us suppose the following conditions:  
\begin{align}
z_{1}^2=\zeta_{+}^2\,,\quad z_{2}^2=\zeta_{-}^2\,. 
\end{align}
%in (\ref{rho-para0}) and (\ref{rho-para1})\,, and then we obtain
Then the parameters in (\ref{rho-para0}) and (\ref{rho-para1}) are expressed as 
\begin{align}\begin{split}
\ck=&\, K\frac{\zeta_{-}^2-\zeta_{+}^2}{(z_{1}^2-z_{2}^2)^2}=-K\frac{1}{z_{1}^2-z_{2}^2}\,,\\
\rho_{11}^{(0)}=&\,\rho_{22}^{(0)}=\rho_{11}^{(1)}=\rho_{22}^{(1)}=\frac{\ck}{2}\,,\\
\rho_{21}^{(0)}=&\,-\ck\,,\qquad \rho_{21}^{(1)}=\frac{z_{1}}{z_{2}}\rho_{21}^{(0)}\,,\\
\rho_{12}^{(0)}=&\,\rho_{12}^{(1)}=0\,.
\end{split}\end{align}
This parametrization corresponds to the limit where the poles and zeros of the twist function (\ref{eq:twist-new}) coincide.
In this case, the resulting action and Lax pair are given by, respectively, 
\begin{align}
S[g_1,g_2]=&\, \bigg\{\sum_{r=1}^{2}\int d^2\sigma \,\frac{k}{2}\tr(J_{r,+},J_{r,-}) +\frac{k}{3}\int \tr( J_1\wedge J_1\wedge j_{1})
-\frac{k}{3}\int \tr( J_2\wedge J_2\wedge J_{2})\no\\
&\;\;\;-k\int d^2\sigma \tr(J_{2,+}^{(0)}J_{1,-}^{(0)}) - \frac{z_{1}}{z_{2}}k\int d^2\sigma \tr(J_{2,+}^{(1)}J_{1,-}^{(1)})\bigg\} \,,
\label{ABL-lambda-action}
\\
\cL=&\,\left( J_{2,+}^{(0)} + z J_{2,+}^{(1)}\right)d\sigma^{+} + \left( J_{1,-}^{(0)} + \frac{z_{1}}{z_{2}z}J_{1,-}^{(1)}\right)d\sigma^{-}\,.
\label{ABL-Lax1}
\end{align}
This is nothing but the Lagrangian and Lax pair of a $G\times G/H$ sigma model related to the tripled $G/H$ $\lambda$-model formulation\footnote{
The deformation parameter $z_{1}/z_{2}$ corresponds to $\tilde{\lambda}$ in \cite{Hoare:2019mcc}.} \cite{Hoare:2019mcc, Levine:2021fof}\,. 
Notably, in the limit $z_{1}/z_{2}\to0$\,, the action (\ref{ABL-lambda-action}) reduces to that of the GMM model \cite{Guadagnini:1987ty}. 

\medskip 

 The flatness condition for this Lax pair is obtained as
 \begin{align}
 0=&\, \pa_{+}\cL_{-}-\pa_{-}\cL_{+}+[\cL_{+},\cL_{-}]\no\\
 =&\,
\pa_{+}j_{1,-}^{(0)}-\pa_{-}j_{2,+}^{(0)} + \left[j_{2,+}^{(0)},j_{1,-}^{(0)}\right] + \frac{z_{1}}{z_{2}} \left[j_{2,+}^{(1)},j_{1,-}^{(1)}\right] 
\no\\
&+
\frac{z_{1}}{z_{2}z}\left(\pa_{+}j_{1,-}^{(1)} +\left[ j_{2,+}^{(0)}, j_{1,-}^{(1)} \right] \right)
+ z\left( -\pa_{-}j_{2,+}^{(1)} + \left[ j_{2,+}^{(1)}, j_{1,-}^{(0)} \right]\right)\,,
\label{ABL-lambda-flatness}
\end{align}
and this is equivalent to the equations of motion of the model. 
Note that as we take the limit $z_{1}/z_{2}\to0$\,, the term with $1/z$ is lost, and thus we cannot reproduce all of the equations of motions.
A possivle way to care this point would be to prepare another Lax pair by scaling the spectral parameter as $z'=zz_{2}/z_{1}$ \cite{Arutyunov:2020sdo}.

\medskip

The background metric and $B$-field for the model (\ref{ABL-lambda-action}) is given by, respectively, 
\begin{align}
\begin{split}
G_{\mu\nu}dx^{\mu}dx^{\nu}=&\,
\ck\sum_{r=1}^{2}\left(d\theta_{r}^2+\sin^2\theta_{r}d\phi_{r}^2\right)
+\ck\left(d\psi + \cos\theta_{1}d\phi_{1} + \cos\theta_{2}d\phi_{2}\right)^2\\
&+\frac{2z_{1}}{z_{2}}\ck\Big[-\sin\theta_{1}\sin\theta_{2}\cos\psi \,d\phi_{1}d\phi_{2} + \cos\psi \,d\theta_{1}d\theta_{2}\\
&\hspace{48pt} + \sin\theta_{1}\sin\psi \,d\phi_{1}d\theta_{2} + \sin\theta_{2}\sin\psi\, d\theta_{1}d\phi_{2} \Big]\,,
\end{split}\\
\begin{split}
B=&\,
\ck\left(d\psi + \cos\theta_{1}d\phi_{1}\right)\wedge \left(d\psi + \cos\theta_{2}d\phi_{2}\right)\\
&+\frac{z_{1}}{z_{2}}\ck
\Big[ -\sin\theta_{1}\sin\theta_{2}\cos\psi \,d\phi_{1}\wedge d\phi_{2} + \cos\psi\, d\theta_{1}\wedge d\theta_{2} \\
&\hspace{42pt}+ \sin\theta_{1}\sin\psi \,d\phi_{1}\wedge d\theta_{2} + \sin\theta_{2}\sin\psi\, d\theta_{1}\wedge d\phi_{2}\Big]\,.
\end{split}
\end{align}
This is a deformed GMM background with a parameter $z_{1}/z_{2}$\,. 
When $z_1/z_2=0$\,, the target space of the original GMM model is reproduced.

\section{Conclusion and Discussion} 

In this paper, we have derived the ABL model from a 4D CS theory with  a meromorphic one-form (\ref{eq:twist-new}) with four double poles 
and six simple zeros by specifying a boundary condition. Then we have explicitly derived the sigma-model background with metric and anti-symmetric two-form 
(i.e., the ABL background). 
As its special cases, we have presented an anisotropic $T^{1,1}$ model and a $G/H$ $\la$-model.
The latter can be regarded as a one-parameter integrable deformation of the GMM model.

\medskip

It would be very interesting to consider the ABL background in the context of AdS/CFT. 
The first task is to find out a possible string-theory embedding of the ABL background. 
It is nice to try to identify the remaining components of type IIB supergravity for the ABL background. 
A possibility is to consider a variant of the ABL model for $G=SL(2)$ and $H=U(1)$.}
This is a natural extension of the work \cite{PandoZayas:2000he}, which considered the GMM model for $G=SU(2)\times SL(2)$ 
and $H=U(1)\times U(1)$ so as to be a supergravity background.
Once the ABL background has been embedded into string theory, 
it would be nice to explore the dual gauge theory. 

\medskip 

Another future direction is to consider a relation between the present result and 6D holomorphic CS theory \cite{Costello:talk,Bittleston:2020hfv}. 
Along this line, it may be possible to derive a new family of 4D integrable systems. We would like to report 
some results in the near future in another place.

\medskip 

We hope that the ABL background we have derived would open up a new arena in the study of the integrability in AdS/CFT.

\subsection*{Acknowledgments}

We would like to thank H.~Y.~Chen and B.~Vicedo for useful discussions 
during the online workshop on ``Online 2020 NTU-Kyoto high energy physics workshop.'' 
The work of O.F.\ was supported by Grant-in-Aid for JSPS Fellows No. 21J22806.
The work of J.S.\ was supported in part by Ministry of Science and Technology (project no. 109-2811-M-002-539), 
National Taiwan University.
The works of K.Y.\ was supported by the Supporting Program for Interaction-based Initiative Team Studies (SPIRITS) from Kyoto University, and JSPS Grant-in-Aid for Scientific Research (B) No.\,18H01214. This work was also supported in part by the JSPS Japan-Russia Research Cooperative Program.

%%%%%%%%%%%%%%%%%%%%%%%%%%%%%%%%%%%%%%%%%%%%%%%%%%%%%%%%%%%%%%

\section*{Appendix}

\appendix

\section{A scaling limit of the ABL model}

It is helpful to give a brief explanation about a scaling limit of the ABL model considered in \cite{Arutyunov:2020sdo}.

\medskip

Let us first redefine the parameters of the model as
\begin{align}
  z_1=1\,,\qquad  z_2=\frac{1}{\alpha}\,,\qquad K=\frac{\la^2_2}{\alpha^2}\,,\qquad \zeta_1=\frac{\la}{\la_2\alpha}\,,\qquad \zeta_2=\frac{\la_1}{\la}\,,\label{limit}
\end{align}
and then take the $\alpha \to 0$ limit while keeping $\la_1\,,\la_2$ and $\la$ fixed.
In this limit, the twist function (\ref{eq:twist-new}) becomes
\begin{align}
   \varphi_1(z)= \frac{2z(\la_1^2-\la^2 z^2)}{(z^2-1)^2}\,.
\end{align}
The double poles and simple zeros of $\varphi_1(z)$ are listed as 
\begin{align}
    \mathfrak{p}=\{\pm 1\}\,,\qquad \mathfrak{z}=\{0, \pm\frac{\la_1}{\la}\}\,.
\end{align}
The Lax pair (\ref{ABL-Lax}) reduces to
\begin{align}
    \cL_+(z)&=\frac{1}{\la^2z^2-\la_1^2}\left((\la^2-\la_1^2)\left(J_{1,+}^{(0)}+z\,J_{1,+}^{(1)}\right)+\la^2(z^2-1)\,J_{2,+}^{(0)}\right)\,,\\
    \cL_-(z)&=J_{1,-}^{(0)}+z^{-1}\,J_{1,-}^{(1)}\,,\label{Lax-lim}
\end{align}
and the  resulting action is given by 
\begin{align}
    S[g_1,g_2]&=\sum_{r=1}^2\int d^2\sigma\,\left(\frac{\la^2}{2}\,\left\langle J_{r,+}^{(0)},J_{r,-}^{(0)}\right\rangle+\frac{\la_r^2}{2}\,\left\langle J_{r,+}^{(1)},J_{r,-}^{(1)}\right\rangle  \right) -\la^2 \left\langle J_{2,+}^{(0)},J_{1,-}^{(0)}\right\rangle\no\\
    &\quad+\la^2\,I_{WZ}[g_1]-\la^2\,I_{WZ}[g_2]\,.\label{action-lim-ABL}
\end{align}
Note here that the flatness condition of the Lax pair (\ref{Lax-lim}) does not describe all the equation of motion of the action (\ref{action-lim-ABL}).
The specific choice $\lambda^2=\lambda_{1}^2=\lambda_{2}^2$ leads to the GMM model \cite{Guadagnini:1987ty,PandoZayas:2000he}.
This choice corresponds to the limit $z_{1}/z_{2}\to0$ in (\ref{ABL-lambda-action}).

\medskip

The other equations of motion comes from the flatness condition of another Lax pair which can be obtained 
by taking a different scaling limit. 
To see this, let us redefine $z$ as $z/\alpha$\,, and then take the same limit (\ref{limit}).
Then, the twist function (\ref{eq:twist-new}) becomes 
\begin{align}
   \varphi_2(z)= -\frac{2(\la^2-\la_2^2 z^2)}{z(z^2-1)^2}\,.
\end{align}
The double poles and simple zeros of $\varphi_2(z)$ are listed as 
\begin{align}
    \mathfrak{p}=\{0,\pm 1\}\,,\qquad \mathfrak{z}=\{\infty, \pm\frac{\la}{\la_2}\}\,. 
\end{align}
The Lax pair (\ref{ABL-Lax}) becomes
\begin{align}
   \tilde{\cL}_+(z)&=J_{2,+}^{(0)}+z\,J_{2,+}^{(1)}\,,\\
     \tilde{\cL}_-(z)&=\frac{1}{\la_2^2z^2-\la^2}\left(\la^2(z^2-1)\,J_{-,1}^{(0)}+(\la_2^2-\la^2)\left(z^2J_{-,2}^{(0)}+z\,J_{-,2}^{(1)}\right)\right)\,.\label{Lax-lim-2}
\end{align}
We can check that the flatness condition of (\ref{Lax-lim-2}) leads to the remaining equations of motion of the action (\ref{action-lim-ABL}).

\end{document}